\documentclass[prl
,twocolumn%
,aps]{revtex4}
\usepackage{natbib}
\usepackage{graphicx}
\begin{document}

\title{Reply to the Comment on 
`Deterministic Single-Photon Source for Distributed Quantum Networking'}
\author{Axel Kuhn}
\author{Markus Hennrich}
\author{Gerhard Rempe}
\affiliation{Max-Planck-Institut f\"{u}r Quantenoptik, Hans-Kopfermann-Str.1, 85748 Garching, Germany}
\date{\today}
\maketitle

A recent comment \cite{KimbleC} addresses the question whether a deterministic and intrinsically reversible single-photon source has been demonstrated in Ref.\,\cite{Kuhn02}. Although the author of \cite{KimbleC} admits that \cite{Kuhn02} is certainly an advance towards these goals, he has four objections:

First, reversibility has not been demonstrated. This is correct. Ref.\,\cite{Kuhn02} only points out that the Raman process employed to produce the photons is intrinsically reversible, in contrast to  most other single-photon generation schemes. 

Second, the stochastic trajectories of the atoms lead to amplitude and phase fluctuations of the emitted light pulse. Amplitude variations, i.e. photon-emission probabilities $P_{emit}$, are discussed in \cite{Kuhn02}, but phase jitter is neglected. This is well justified, since the velocity of the atoms along the cavity axis is restricted to $\pm 5\,$mm/s, leading to a phase jitter below $\pm \pi/40$ for a $2\rm\mu s$-long pulse. The phase might vary from pulse to pulse, but this has no influence on the Raman process, i.e. reversibility is not affected. A phase-preserving teleportation of atomic superposition states requires a more elaborate technique \cite{Gheri98} which is beyond the scope of \cite{Kuhn02}.

\begin{figure}[t]
	\begin{center}
		\includegraphics[width=1.0\columnwidth]{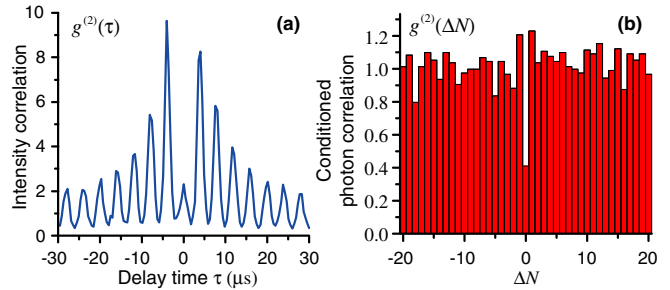}
	\end{center}
	\caption{Correlation function for a flux of 10 atoms/ms. (a) The intensity correlation, $g^{(2)}(\tau)$, oscillates around 1 due to the pulsed excitation of the atoms. For $|\tau|>20\mu s$ and $\tau=0$ the signal is caused by correlations between photons from different atoms. (b) Sub-Poissonian photon statistics in the conditioned photon correlation, $g^{(2)}{(\Delta N)}$. The finite atom-cavity transit time causes the peaks at $\Delta N=\pm 1$.
	}
	\label{fig:highAF}
\end{figure}

Third, it is claimed that atom-number fluctuations would always lead to a photon statistics with $g^{(2)}(\tau)\ge 1$ even if background noise were eliminated. A measurement with increased atom flux and, hence, reduced background contribution shows that this is not the case. Fig.\ref{fig:highAF}\,(a) illustrates that all minima of $g^{(2)}(\tau)$ are below 1. This effect was not visible in \cite{Kuhn02} due to the smaller atom flux. The minima of $g^{(2)}(\tau)$ are due to the fact that the light emission consists of a sequence of bright and dark intervals determined by periodically turning on and off the pump laser. Hence, the photon-emission probabilities of all atoms oscillate in phase, so that no correlations occur between bright and dark intervals. This effect is absent in case of the continuous excitation scheme discussed in \cite{KimbleC}, where out-of-phase Rabi oscillations lead to $g^{(2)}(\tau)\ge 1$. Such a situation cannot be compared to the case realized in \cite{Kuhn02}. 

Forth, and most important, the author of \cite{KimbleC} says that the measured correlation function does not support the deterministic generation of single photons. It is argued that without a-priori knowledge about the presence of an atom in the cavity, the Poissonian atom statistics is simply mapped to the photon statistics. In fact, conditional detection is necessary to obtain a sub-Poissonian photon statistics. In \cite{Kuhn02}, it is therefore {\em`emphasized, that the detection of a first photon signals the presence of an atom \ldots photons emitted during subsequent pump pulses dominate the photon statistics and give rise to antibunching'}. To illustrate this statement further, Fig.\ref{fig:highAF}\,(b) shows the photon correlation, $g^{(2)}(\Delta N)$, after conditioning on the presence of an atom. Here, for every photon detected during a pump pulse, only events during the neighboring bright interval are considered. All these intervals are chained together and then used to calculate $g^{(2)}(\Delta N)$ in the usual way, with $\Delta N$ denoting the difference between interval numbers within this chain. One obtains $g^{(2)}(\Delta N=0)=0.41(6)$ and $g^{(2)}(\Delta N \neq 0)=1.00(9)$, i.e. the photon statistics conditioned on the presence of an atom is sub-Poissonian. For the data recorded with reduced atom flux published in \cite{Kuhn02}, an even smaller value $g^{(2)}(\Delta N=0)=0.25(11)$ is observed.

In summary, the proposed reversal of the photon emission using either the same atom or a second atom in another cavity seems feasible with the current setup, since it allows the triggered emission of several single-photon pulses of negligible phase jitter within a short time interval.

\end{document}